ANALYSIS AND ASSEMBLING OF NETWORK STRUCTURE IN MUTUALISTIC SYSTEMS

Diego Medan[1,6] (corresponding author), Roberto P. J. Perazzo[2], Mariano Devoto[1], Enrique Burgos[3,6], Martín G. Zimmermann[4], Horacio Ceva[3], & Ana M. Delbue[5]

[1] Cátedra de Botánica, Facultad de Agronomía de la Universidad de Buenos Aires, Av. San Martín 4454, 1417DSE Buenos Aires, Argentina. diemedan@agro.uba.ar

[2] Departamento de Investigación y Desarrollo, Instituto Tecnológico de Buenos Aires, Av. E. Madero 399, 1106 Buenos Aires, Argentina.

[3] Departamento de Física, Comisión Nacional de Energía Atómica, Av. del Libertador 8250, 1429 Buenos Aires, Argentina.

[4] Departamento de Física, Facultad de Ciencias Exactas y Naturales de la Universidad de Buenos Aires, Ciudad Universitaria, 1428 Buenos Aires, Argentina.

[5] Departamento de Economía, Facultad de Ciencias Sociales y Económicas de la Universidad Católica Argentina, Av. A. Moreau de Justo 1400, 1107 Buenos Aires, Argentina.

[6] Consejo Nacional de Investigaciones Científicas y Técnicas, Av. Rivadavia 1917, C1033AAJ Buenos Aires, Argentina.

**Abstract**

It has been observed that mutualistic bipartite networks have a nested structure of interactions. In addition, the degree distributions associated with the two guilds involved in such networks (e.g. plants & pollinators or plants & seed dispersers) approximately follow a truncated power law. We show that nestedness and truncated power law distributions are intimately linked, and that any biological reasons for such truncation are superimposed to finite size effects . We further explore the internal organization of bipartite networks by developing a self-organizing network model (SNM) that reproduces empirical observations of pollination systems of widely different sizes. Since the only inputs to the SNM are numbers of plant and animal species, and their interactions (i.e., no data on local abundance of the interacting species are needed), we suggest that the well-known association between species frequency of interaction and species degree is a consequence rather than a cause, of the observed network structure.

**1. Introduction**

The co-existence of plants and animals on Earth has given rise to interactions of the most variegated type. A particularly important type is that of mutualistic interactions, in which species may play an important role on each other. For example, birds feed from fruits while they disperse the seeds; insects feed from the nectar of flowers pollinating them in the process (Herrera and Pellmyr, 2002).

A great amount of research has been devoted to the study of mutualism as a community-level phenomenon (Waser and Ollerton, 2006). In traditional studies the interaction of all active plant and animal species is recorded within a restricted geographical extension (Medan et al., 2002; Memmott et al., 2004). A standard graphical description of these systems can be made through bipartite networks in which nodes (species) are linked (interact) only with nodes of the opposite guild (plants with animals and vice versa) (Albert and Barabási, 2002). These networks may also be mathematically depicted by an adjacency matrix where rows and columns represent the two guilds of species, and a 1 in the intersection of a row and a column indicates that the corresponding species interact, and alternatively a 0 indicates they do not. The network-level pattern of interactions among the mutualist species can also be described through the degree distribution, i.e., the number of nodes for every degree value (number of links). In graph theory, this distribution plays a key role because it provides valuable hints about the internal structure of the network and hence, about some behavioral pattern of the species represented by its nodes (Albert and Barabási, 2002).

An additional tool widely used to describe the organization of mutualist networks was developed by Atmar and Patterson (1993) who originally conceived it to describe how an array of species were distributed in a set of islands of different sizes. They found a pattern in which bigger islands hosted a larger number of species and in turn, smaller islands hosted fewer species but in such a way that the species found in the smaller islands represented a subset of those species found in larger ones. Such pattern was thus defined as "nested" and is represented by an approximately triangular shape of the distribution of 1's in the adjacency matrix, provided that rows and columns have been reordered by increasing number of links. In addition, perfect nestedness is achieved when all 1's are within a region of the matrix delimited by an *extinction curve* or *isocline of perfect nestedness* (hereafter: IPN).

When analyzed using these tools real mutualistic networks display a remarkable degree of internal organization. On the one hand, the adjacency matrix displays nestedness (Petanidou and Ellis, 1993; Bascompte et al., 2003). In a nested matrix, 'generalist' species (those with a high degree) interact with their generalist counterparts constituting a highly-connected 'core' within the matrix. In addition, 'specialists' (species with low degree) also tend to interact with generalists, rather than among themselves. On the other hand, most observed degree distributions of mutualistic networks fit a truncated power-law (hereafter: TPL) function (Jordano et al., 2003). Both features indicate that mutualistic networks are far from being a random collection of species and interactions. If this were the case all the 1's of the contact matrix should appear randomly distributed and the degree distribution should fit a Poisson distribution.

Both power-law degree distributions and nestedness have been associated with a high tolerance of pollination networks to species extinctions (Memmott et al., 2004; Vázquez, 2005) and unexpected perturbations (Bascompte et al., 2003). However, the underlying causes of both features and the reasons why they are so widespread have not yet been properly established. Two explanatory approaches exist. First, Jordano et al. (2003) suggest preferential attachment (i.e. the tendency of new species to interact with already highly connected partners) as the main mechanism (see also Guimarães et al. 2005) for network creation. Additional factors preventing particular interactions such as phenological incompatibilities represented by forbidden links (hereafter: FL) are mentioned (Waser et al., 1996; Jordano et al., 2003) also. A second approach (Vázquez and Aizen, 2003, 2004, 2006; Vázquez, 2005) suggests that species degree may simply be a function of species abundance in the community. Species represented by many individuals will have the most links, and rare species will have the fewest. This approach is based on a model of interaction matrices which attempts to reproduce the degree distribution of real networks. However some questions remain that call for a unified picture for the organization of mutualist networks. First, concerning the relationship between degree distributions and local species abundance, the reverse interpretation is also possible, i.e., that accumulation of interactions occurs first, and the ensuing higher reproductive success leads to local abundance. Second, remarkably, nestedness and power-law patterns (including truncated variants) always co-occur suggesting that both features are, in fact, "two sides of the same coin". The claim that nestedness and TPL can be linked to a mechanism of preferential attachment is hard to work out. That mechanism has been originally developed by Barabási and Albert (1999) in a model involving a purely stochastic attachment rule of newly added nodes to a continuously growing network. This framework does not seem appropriate for real mutualist networks where it is necessary to preserve a given number of nodes and links. Third, although the IPN represents an intuitively sound and apparently straightforward approximation to the limiting shape of a perfectly nested distribution, on closer view some caveats arise. The curve is said to be a function of the number of islands, the number of species and the probability of contact between them, but its derivation arises from a purely geometric interpolation and is not clearly linked to any underlying biological mechanism or statistical analysis.

In the present paper we provide a unified picture for the internal organization of mutualist networks deriving the empirical evidence of nestedness, degree distributions and the theoretical IPN, from a few biological hypotheses concerning the interaction of mutualist species. The structure of the paper is as follows:

1) We provide a mathematical derivation of the IPN and also analytically demonstrate that, as long as nestedness is measured by comparing the pattern of contacts to that curve, a network with a nested pattern of interaction will present at the same time a truncated power law for the degree distribution, and vice versa.

2) We present some theoretical implications of this derivation. We show how the cumulative degree distributions can be used to derive a direct measure of the IPN. We also show why the power laws that are experimentally observed appear to be truncated and that the degree distributions of either plant and animal species of a given mutualist system remain mathematically related to each other.

3) Finally, we show how the collective nested pattern of contacts and TPL degree distributions can naturally emerge from a Self-organizing Network Model (SNM), which is first presented here, for the allocation of interactions between species within a mutualistic bipartite network. The SNM model thus provides a biologically plausible foundation for the IPN and is validated by the quantitative fitting of degree distributions empirically observed in systems of widely different sizes. We also show how the SNM can be used to obtain a quantitative estimate of the effects of forbidden links on the behavior of the network.

## 2. The isocline of perfect nestedness

### 2.1 The analytic expression

As mentioned above, a nested pattern of contacts is represented by an approximately triangular shape of the distribution of 1's in the adjacency matrix, and perfect nestedness is achieved when all 1's are within a region of the matrix delimited by the IPN. By quantifying the departure from this perfectly nested theoretical matrix, Atmar and Patterson (1993) proposed a measure of the disorder of a real network, expressed as a temperature, which has been widely used both in biogeography and community ecology (Memmott et al., 2004; Bascompte et al., 2003; Jordano et al., 2006). The curve is said to be a function of the number of islands, the number of species and the probability of contact between them. If $n$ and $m$ are respectively the number of columns and rows of the matrix and $\phi$ is the probability of a contact between both types of species, a perfectly ordered matrix is expected to concentrate the $nm\phi$ 1's in a region similar to the one limited by the two straight lines UW1 and UW2 shown in Fig. 1, where U is the point of coordinates $(n\phi, m\phi)$.

The IPN is however not expected to involve straight lines or vertices. It is therefore represented by a function that is a continuous modification of the straight segments. Neither the derivation of the IPN nor its analytic expression are provided by Atmar and Patterson (1993). We develop below an analytic expression of an IPN that has all the same properties as that proposed by Atmar and Patterson (1993) although, in absence of a formal expression for the latter, the equivalence is only supported by numerical evidence.

We describe the IPN in terms of the two continuous variables $a$ $(0 \leq a \leq n)$ and $p$ $(0 \leq p \leq m)$ that can be assimilated respectively to the columns (animal species) and rows (plant species) of the adjacency matrix. This approximation may be considered to be exact in the limit of very large systems.

It is convenient to consider separately the two branches of the IPN, lying respectively below (branch 1) and above (branch 2) the diagonal of the adjacency matrix (see Fig.1).

To begin with, we write the coordinates *(a', p')* of each point of the segment UW1 as $a' = \beta + d \cos\theta = \beta + d\, n/D$, and $p' = d \sin\theta = d\, m/D$, where the segment d is parallel to the main diagonal, $D = (n^2 + m^2)^{1/2}$, and $\beta$ is a parameter $(0 < \beta < n)$. Next, we map each point *(a', p')* of the segment UW1 into the corresponding point *(a₁, p₁)* on branch-1. We do this by means of a continuous stretching or shortening of each segment *d*. More specifically, this amounts to multiplying *d* by a factor $(d/d_0)^\mu$, where $d_0$ is the *d*-segment associated with the crossing point of branch-1 and UW1. Hence, using $d/\phi D = (n - \beta)/n$, the Cartesian coordinates $a_1$ and $p_1$ of the points of branch-1, are:

$$a_1 = \beta + \frac{n}{D} d [d/d_0]^\mu = \beta + \phi(n-\beta)\left[\phi D \frac{n-\beta}{n d_0}\right]^\mu \tag{1}$$

$$p_1 = \frac{m}{D} d(d/d_0)^\mu = m\phi \frac{(n-\beta)}{n}\left[\phi D \frac{n-\beta}{n d_0}\right]^\mu \tag{2}$$

A completely analogous set of equations can be found for branch-2. We use for this case the parameter $\eta$ $(0 < \eta < m)$ playing a role analogous to $\beta$ in the previous expression:

$$a_2 = \frac{n}{D} d[d/d_0]^\mu = n\phi \frac{(m-\eta)}{m}\left[\phi D \frac{m-\eta}{m d_0}\right]^\mu \tag{3}$$

$$p_2 = \eta + \frac{m}{D} d[d/d_0]^\mu = \eta + \phi(m-\eta)\left[\phi D \frac{m-\eta}{m d_0}\right]^\mu \tag{4}$$

The constants $d_0$ and $\mu$ are determined by imposing that the area limited by the isocline is $nm\phi$ and that both branches of the curve meet at the diagonal with a continuous first derivative. The two conditions correspond to:

$$1 = \phi(\mu+1)(\mu+2) \tag{5}$$

$$d_0^\mu = (\phi D)^\mu 2\phi(\mu+1) \tag{6}$$

that completely specify the IPN. We have thus built up a set of two of parametric equations, one for each branch, with parameters $\beta$ and $\eta$, respectively.

### 2.2 The cumulative degree distribution

The IPN presented above contains also the information of the two degree distributions for rows and columns. This is so because it provides the number of contacts of each species with its mutualist counterparts. In graph theory parlance this is the degree of each node of the bipartite graph. One further step is needed to link the IPN to the degree distribution because the latter measures *how many* animal or plant species *have the same degree* (in what follows for shortness we will omit the word species, and refer to plants and animals). To trace the relationship of the IPN with the degree distributions of plants or animals (rows and columns), we write it as $p = p(a)$. We assume that all plants $p$ and animals $a$ have been ordered in such a way that $p(a)$ is a monotonously decreasing function.

A different way of reading this curve is by realizing that a point $(a_o, p_o)$ (see Fig. 2) is directly related to the degree of animals and plants. Indeed the value $p_o$ indicates the degree of the animal $a_o$ and vice versa, $a_o$ indicates the degree of the plant $p_o$. Since the curve is monotonous, $a_o$ is also the number of animals that are connected to $p_o$ *or more* plants (shaded area in Fig. 2) and, equivalently, $p_o$ is the number of plants that are connected to $a_o$ *or more* animals (shaded area in Fig. 2). This means that the IPN can also be read as a cumulative degree distribution.

These cumulative distributions are reported in the literature with no reference to the IPN, and are usually normalized to 1. In Fig. 3 we show an example of the direct comparison of the IPN with the two possible cumulative degree distributions, one for plants and the other for animals. The example is taken from a real system as reported by Robertson (1929). In order to compare them with the IPN the two cumulative degree distributions are not normalized as they are usually shown in the literature; in addition the degree distribution of plants must be read as referred to the vertical axis, while that of animals must be referred to the horizontal axis. The endpoint at $a \approx 300; p \approx 1$ (indicated with an arrow in Fig. 3) shows that there are no plants that are connected to more than 300 animals. Analogously, the endpoint at $a \approx 1; p \approx 240$ indicates that there are no animals with a degree greater than 240. In a perfectly ordered system, i.e. one with vanishing temperature, both degree distributions would have reached the two corners of the matrix, namely $a = 456; p = 1$ and $a = 1; p = 1428$.

### 2.3 The degree distributions

From the above arguments one can readily see that the regular degree distributions can be related to the two possible derivatives of the IPN, either $dp(a)/da$ or $da(p)/dp$ where $a(p)$ is the inverse function of $p(a)$. Alternatively this can be seen by approximating the IPN by a stair-like function obtained by dividing the $a$-axis into equal bins of a width $\Delta a$ (see Fig. 2). The IPN may thus be replaced by a stair-like line in which all steps have the same width $\Delta a$ and a varying height. Within this approximation all the plants belonging to the same step of the stair are $\Delta p$ in number, and have the same degree that is equal to the value $a_o$ that is at the center of the interval $\Delta a$. Since $\Delta p \approx \Delta a |dp(a)/da|$ it follows that, in the limit in which $\Delta a \approx 1 \ll a_{max}$, the stair-like curve approaches the IPN and the (ordinary) degree distribution of the animals can well be approximated by the derivative $|dp(a)/da|$ as anticipated above. The absolute value is used to ensure that the degree distribution is a positive number. A completely similar argument can be made for plants, reading the IPN as $a = a(p)$ and approximating it by a stair-like curve with steps of equal height $\Delta p$ and varying widths $\Delta a$.

In the left panel of Fig. 4 we show as an example the log-log plot of the analytic expressions of both degree distributions for an experimentally observed adjacency matrix (Robertson, 1929). As can readily be seen both have the shape of truncated power laws. Within the present derivations it is not necessary to resort to forbidden links or other biological justification to explain the truncation of the power law of the degree distribution.

### 2.4 Consistency of the degree distributions for plants and animals

The two degree distributions for rows and columns are related to the two possible derivatives of the IPN, namely one in which $a$ is the independent variable and the other in which $p$ plays that role. A way to relate the derivative $dp(a)/da$ with that of the inverse function $da(p)/dp$ is by realizing that both functions are related to each other in the same way as branch-1 (in Eqs. (1,2)) and branch-2 (in Eqs. (3,4)). In fact the function $a(p)$ can be mapped into $p(a)$ in the same way. This can be made by setting $\eta/m = \beta/n$ and realizing that then $a_1; p_1; a_2$ and $p_2$ fulfill:

$$ma_2 = np_1 \quad ; \quad np_2 = ma_1 \qquad (7)$$

By choosing first $a$ as the independent variable and next $p$, the degree distributions are

$$\frac{dp_1}{da_1} = \frac{dp_1/d\beta}{da_1/d\beta} \qquad (8)$$

$$\frac{da_2}{dp_2} = \frac{da_2/d\eta}{dp_2/d\eta} \qquad (9)$$

and, by using Eq.(7), both remain related as

$$\frac{da_2}{dp_2} = \left(\frac{n}{m}\right)^2 \frac{dp_1}{da_1} \quad ; \quad \frac{da_1}{dp_1} = \left(\frac{n}{m}\right)^2 \frac{dp_2}{da_2} \qquad (10)$$

These equations indicate that if both distributions are plotted as usual in a log-log plot (using in each case the appropriate independent variable), they show the same slope, thus approaching a power law with the same exponent $\nu$. According to the above equations both curves can be made to collapse into each other. This requires stretching the x-axis of one of the degree distributions by a factor equal to the ratio of the two dimensions of the adjacency matrix (as suggested by Eq. (7)), and dividing the resulting distribution by the square of the same factor as indicated in Eq. (10). This mapping is exactly fulfilled by the analytic curves but is only approximately fulfilled by the empirical data. In Fig. 4 we show the experimental degree distributions of rows and columns of the Robertson (1929) matrix together with the distributions obtained from the derivatives of the IPN. The minor departures between the empirical data and the theoretical curves may arise from the fact that the real system does not correspond to a perfectly nested bipartite network. The agreement should improve for systems with a greater nestedness.

In the right panel we show how the two degree distributions collapse into each other after the renormalization of the column data following the calculations explained above. The two analytic curves are exactly superimposed. The empirical data of rows and columns show similar slopes as predicted by the theory. Despite the fact that both degree distributions show sizable fluctuations for large degrees, they are quite consistent with each other once the renormalization procedure is carried out.

The exponent $\nu$ of the power law associated with the degree distributions can also be discussed with the aid of Eqs. (2) through (6). In the limit $a/n \to 0$ the degree distribution approaches the power law $(a/n)^{-\mu/(\mu+1)}$. On the other hand from Eq. (6) it follows that $\mu$ is the positive root of the equation $1 = \phi(\mu + 1)(\mu + 2)$. We therefore conclude that $\nu = -\mu/(\mu+1)$ is a function of the probability of contacts between mutualist counterparts and is therefore independent of the number of species involved in the system.
If expanded for small values of $\phi$ then $\nu = -1 - \sqrt{\phi} + O(\phi)$.

## 3. A model for mutualist bipartite networks

The two distinct features of mutualist networks, namely their nestedness and the truncated power law for the degree distributions, have been hitherto considered in the literature to be independent features. We have shown that as long as a nested distribution of contacts is one that approaches the IPN, they are actually two ways of looking at the same structure and they therefore should not be considered unrelated. In fact both are intimately related and the same biological arguments should be applied to understanding both features.

### *3.1 The Self-organizing Network Model (SNM)*

We now turn to considering the biological mechanisms that may cause a mutualistic system to approach the nested pattern of contacts that we have just discussed. From the point of view of the theory of complex systems (Bar-Yam, 1997) a proper understanding can be achieved if the global features of a system can be explained in terms of a minimal set of microscopic interactions among its constituents. The model that we present below has precisely the aim of accounting for the main features of mutualist networks in terms of the structure of interactions between species. The SNM has to meet several basic requirements. In the first place it should represent a gradual ordering process so as to describe the partially ordered situations that are found in real world mutualist systems. Secondly it should produce data amenable to the same calculations that are used to study such systems, namely to determine its temperature (Atmar and Patterson 1995) or its cumulative degree distributions. Finally the rules applied for the gradual ordering of the system must approach asymptotically to a perfectly nested organization. If this is achieved a side benefit of the model would be its use as a benchmark to test other hypotheses that have been suggested to play a relevant role in the organization of mutualistic webs, e.g. the presence of forbidden links (FL).

*Concept and aim.* The SNM is an iterative procedure in which the only inputs are the two dimensions of the adjacency matrix (i.e. the number of plants and flower visitors) and the total number of contacts. The model is inspired in evolutionary computational techniques in which species are assumed to change progressively the pattern of contacts. At the start, contacts are randomly distributed among mutualists, as if all information on network wiring had been suddenly lost in a real-world mutualistic web. During the subsequent iterative procedure the SNM reconstructs the web structure by applying a simple rule.

*Assumptions.* The following assumptions are made: (1) In order to facilitate the comparison with real systems, the initial number of species and interactions is kept fixed; i.e. no extinctions among species are allowed, and interactions are allowed only to reallocate between species. (2) The proportion of FL, i.e. impossible interactions, such as those caused by morphological or phenological barriers, does not change with time.

*Basic operation.* Initially FLs are randomly assigned to fixed positions in the matrix, and the interactions are randomly distributed in adjacency matrix, with the only restriction that all mutualists have at least one interaction. Interactions and no-interactions among species are coded respectively with a 1 and a 0. In each iteration of the SNM the following two steps are alternatively applied to the rows and columns of the adjacency matrix: (a) two sites of a column (row) - species respectively having a 1 and a 0 are randomly selected, (b) the 1 is swapped with the 0 if a specified acceptance rule it satisfied (see below), otherwise the swap is rejected.

*Swap acceptance rule.* The swap is only accepted if it satisfies three conditions: (i) The degree of the new partner is higher than that of the previous partner. This rule favors the allocation of interactions to already well-connected species, i.e., it encourages interaction with generalists, (ii) The swap can not take place if any species loses its last interactioni.e. if there is an extinction. (iii) The swapped interaction can not be allocated to a FL position. This swap acceptance criterion may be regarded as a modification of the usual rule of preferential attachment (see below).

*Stopping criteria.* We have used two criteria to stop iterating the model. The first one involves a comparison between the level of nestedness of the model and that of some real-world web. A useful measure of nestedness is the above-discussed Atmar and Patterson (1993) 'temperature'. Temperature provides an estimate of the degree of disorder of the matrix or, equivalently, its departure from the maximum possible nestedness. We calculated it at selected stages of any run of the SNM using the Nestedness Calculator (Atmar and Patterson, 1995) (see Bascompte and Jordano, 2005 and Jordano et al., 2006) for a similar application). Whenever the temperature of the model ($T_m$) closely matches that of the real system ($T_r$), the simulation ends. The second stopping criterion focuses on degree distributions: the simulation ends whenever a good qualitative 'match' between the degree distributions of the SNM run and of the real web is obtained.

*Biological plausibility.* The reallocation of contacts promotes the interaction with generalists. Such change can be assumed to simulate either the adaptation or the replacement of an existing species by another one that makes a more efficient use of the available contacts. Although this change entails increased competition, the interaction with a 'popular' mutualist may be preferable for two reasons: a generalist species offers 'guaranteed' efficacy (were this not the case, it would not have so many partners), and it is dependable in the long run (having many mutualists, its own survival is assured). Running the SNM using a contrary rule (i.e. swapping the available contact with a partner species having a lesser degree, an alternative that minimizes competition with other species of the same guild), results in all species tending to share evenly all the available contacts (results not shown). In a sense, one could say that all species tend to become equally specialized because all tend to have the same number of counterparts.

*SNM and preferential attachment.* The swap acceptance criterion used in the iterative procedure favors the progressive allocation of interactions with already well-connected species and also species of a each guild tend to become generalists by acquiring the largest possible number of available contacts. From this point of view this criterion is reminiscent of preferential attachment. Barabási and Albert (1999) introduced a stochastic growth model, in which new nodes are added continuously and attach themselves to existing nodes, with probability proportional to the degree of the target node (known as 'preferential attachment rule'). At variance with this model, in the SNM the topology of a non-growing network with a fixed number of nodes is progressively reshaped: in each iteration a connection between two nodes of a different kind is rewired to favor a contact with a more connected node.

### 3.2 Consistency of the SNM with perfect nestedness

We first checked the consistency of the SNM with the theoretical IPN. In Fig. 5 we show the frequency distributions of the degree of animal species of the Kato et al. (1990) system in three stages of the SNM with no FL and using the swapping criterion that promotes increasing generalization. Notice that an approximate TPL develops without the need of introducing FL. For comparison we also show in the inset of Fig. 1 the same frequency distributions for the same system but arising when the alternative swap criterion promoting interactions with lower-degree mutualists (increasing specialization). The frequency distribution of contacts is seen to approach the unrealistic situation in which all species tend to have the same number of contacts.

A further check is made in Fig. 6 in which the internal ordering of the system is gauged using the change in the temperature parameter $T_m$ with an increasing number of iterations of the SNM in which the swapping criterion promoting generalization is used. $T_m$ is seen to tend to 0, proving that the SNM converges to an IPN that is indistinguishable from that of Atmar and Partterson (1995) . In the case in which the alternative swapping criterion is used, $T_m$ tends to grow (data not shown), and therefore a nested matrix is never approached.

The gradual changes in the adjacency matrix of the network of the same system are shown in Fig. 7 in several snapshots of the gradual reordering produced by the SNM. The rows and columns of the adjacency matrix are ordered from left to right and from bottom to top in a decreasing number of contacts (rows and columns with equal number of contacts are randomly ordered). The stages of ordering are identified with the same letters that appear in Fig. 2. The experimental adjacency matrix is also compared with an intermediate stage of the SNM. The asymptotic "perfect" ordering that is reached after a very large number of iterations is also shown.

### 3.3 A comparison with real world systems

The quality of the model is tested against five real-world pollination networks, which span the full size range known for this type of webs (Table 1). This comparison can easily be made because the proportion of FL and the total number of iterations are the only free parameters of the model. In all cases, the SNM yields degree distributions and levels of nestedness that agree with the experimental observations.

Theoretical and experimental cumulative degree distributions are shown in Fig. 4. For simplicity, given that the analyses based on plants yielded similar results, we only show degree distributions of animals (corresponding to columns of the adjacency matrix). In all cases a highly satisfactory agreement is found. The two smaller real systems (Dupont et al., 2003; Devoto et al., 2005) approximately fit an exponential distribution. For the three larger systems (Kato et al., 1990; Clements and Long, 1923; and Robertson, 1929) the SNM generates degree distributions which approach TPLs.

A good agreement is found using either stopping criteria for the SNM. The temperature is a statistical parameter that is less stringent than a direct comparison of the degree distribution curves. Therefore it should not be surprising to find SNM stages with different degree distributions to have the same temperature. Likewise, with the nestedness stopping criterion, the experimental power-law portions of the distributions are consistently larger than the theoretical ones (Fig. 4), indicating that for $T_r \approx T_m$ the SNM provides less nested systems. The use of the second stopping criterion improves considerably the fit of the degree distributions (see Table 1).

Remarkably, the behavior of the SNM model does not change essentially when the proportion of FL changes. Approximate TPL distributions are found even in the absence of FL. If the proportion of FL is increased the only effect is to reduce the speed of the self organization process. In the limit in which the proportion of FL is very large the network hardly evolves as there are practically no available pairs to swap and so the degree distribution does not develop TPL features. On the other hand nestedness is not affected by the presence of an intermediate proportion of FL, the only change being the length of the power law regime. In the panel devoted to the Kato et al. (1990) system in Fig. 4 we compare the results obtained with 0% and 80% of FL. In the latter case the number of iterations (greater than 500,000) is large enough to reach the asymptotically ordered state. These figures indicate that for a broad range of proportions of FL, the results are the same as running the SNM without FL and a smaller number of iterations.

**4. Discussion and conclusions**

We investigated the IPN and showed some examples from real systems. We proved that the cumulative degree distributions provide a direct measure of such a curve when they are not normalized as they usully are when presented in the literature. The degree distributions for rows and columns can therefore be related to the derivatives of the IPN. We also proved that the degree distributions of a same adjacency matrix are closely related to each other. In fact a very simple geometric trick can be used to map one set of data onto the other, thus providing a check for the consistency of empirical observations.

We also proved mathematically that for a perfectly nested organization, both the degree distributions and the IPN approach a truncated power law. In addition, the power law exponent can be obtained in this limit and we proved that it depends only on the probability of contacts between the mutualist partners. This is a universal property that allows a direct comparison between different systems. These considerations help to understand the widely observed truncation of the power law adjusting the degree distribution. However, these proofs largely depend on the assumption of large systems. Real mutualist networks are rather small and most statistical features remain obscured by this fact. Empirically observed distributions that are usually claimed to be adjusted by a power law hardly contain more than a few tens of species before truncation becomes a dominant feature. Under these conditions, truncation may be the result, among other possible factors, of finite size effects.

We have next introduced a SNM that explains the nested organization pattern in terms of the kind of interaction between the mutualist partners. On the one hand, using the SNM we were able to generate networks that fit very well the degree distribution and level of nestedness that are observed in systems of the most diverse sizes. On the other hand, the SNM is fully consistent with theoretical patterns of perfect nested order that have been previously introduced, thus providing a plausible biological explanation for this widely observed property. We believe we have made progress in understanding the interaction behavior of mutualist species by identifying the relevant element primarily responsible for the global organization of the network.

Several aspects of the SNM deserve commentary: first, we have shown that contacting preferentially to already highly connected partners drives the system towards a nested organization. This feature reminds the rule of preferential attachment that has a widespread application in network theory. However, that model can not be applied to the case of mutualist networks without strong conceptual changes. The SNM could be regarded as the closest one can get to preferential attachment whe modelling a bipartite network with constant number of nodes and links.

Secondly, as the model assigns all species an equal opportunity to interact, the fact that a given species accumulates many interactions is not a consequence of its being more abundant and/or more frequently interacting than other species in the system, but of the self-organization process itself.

Thirdly, a better match of the degree distributions requires that the SNM be iterated beyond the point where $T_r \approx T_m$ at least for larger systems. Although variants of the SNM could perhaps be developed to correct this, it is important to bear in mind that nestedness being a statistical feature of the system, in general one can not expect to measure it with a single parameter such as $T_m$. This is essentially the second moment of the statistical distribution of contacts with respect to the IPN and one should therefore expect that different distributions may be associated with the same temperature.

Fourthly, the SNM is robust against the introduction of FL. An important consequence is that, even if FL were present in very high density, this should not be an obstacle for the development of nestedness. On the other hand, truncated degree distributions emerge also in the absence of FL. Therefore the results of the SNM suggest that the presence or absence of FL in a real mutualistic network cannot be established from the observation of the statistical features of the web.

In conclusion, given the number of plants, animals, and interactions, the model generates a highly realistic pattern of interactions. It shows *in which proportion* species will behave as extreme generalists, moderate generalists, and extreme specialists. Many biological factors are probably involved in the establishment of every particular interaction in each real-world web, and this complexity cannot be accounted for by a model primarily designed to elucidate network structure. Admittedly, the model makes thus no statement about *which* particular species will adopt which role. Explaining the interaction behavior of individual species certainly calls for a much more elaborate model and for the inclusion of additional parameters. Interestingly, one such possible factor (species' frequency of interaction) has been shown to be strongly and positively associated to species degree (Vázquez and Aizen, 2006) which suggests that locally abundant species are prone to accumulate interactions (and rare species, conversely; Stang et. al., 2006), thus being the 'natural candidates' to become the generalists (specialists) in the system. However, since the SNM generates a realistic distribution of species degree under no assumptions on species abundance, the reverse interpretation is also possible, i.e., that accumulation of interactions occurs first, and that the resulting higher reproductive success leads later to local abundance.

Plant-pollinator and plant-disperser networks are complex systems whose structure we are still beginning to understand. Different modelling approaches (e.g. based on some set of individual encounter rules and data on local abundance of mutualist species) seem worth exploring and would perhaps be complementary to the present contribution.

**Acknowledgements**
We thank J. Memmott for providing us the Robertson's 1929 interaction dataset, and two anonymous reviewers for comments that helped to substantially improve previous drafts. This work was funded with a grant from the Agencia Nacional de Promoción Científica y Tecnológica of Argentina (PICT 25450). D.M. and M.G.Z. are associated to CONICET, Argentina.

**TABLE 1**. Main attributes of the five real pollination networks whose parameters were used as input for modelling.

|  | Dupont et al. (2003) | Devoto et al. (2005) | Clements & Long (1923) | Kato et al. (1990) | Robertson (1929) |
|---|---|---|---|---|---|
| Nr. of plant species (P) | 11 | 29 | 96 | 91 | 456 |
| Nr. of animal species (A) | 38 | 101 | 275 | 679 | 1428 |
| Nr. of interactions (I) | 106 | 146 | 923 | 1206 | 15254 |
| Connectance [=I/P*A] | 0.253 | 0.049 | 0.034 | 0.019 | 0.023 |
| $T_r$ | 33.89 | 5.56 | 2.41 | 0.95 | 0.79 |
| $T_m$ | 32.2 | 5.28 | 2.41 | 0.95 | 0.78 |
| Nr. of iterations for a $T_m \approx T_r$ (FL=0) | 60 | 600 | 2,200 | 6,700 | 72,000 |
| Nr. of iterations for a better fit (FL=0) | 60 | 600 | 10,000 | 20,000 | 400,000 |
| Nr. of iterations for a $T_m \approx 0$ (FL=0) | 2,000 | 30,000 | 50,000 | 150,000 | 750,000 |

**Captions to figures**

**FIG. 1**: A matrix of *n* columns and *m* rows is shown together with the IPN that corresponds to a probability of contact $\phi$ between the two mutualistic species. The area limited by each branch of the curve is $nm\phi/2$ and it is the same as either of that the two triangles $T_1 \equiv OUW_1$ and $T_2 \equiv OUW_2$. The IPN is a smooth distortion of the two straight lines $UW_1$ and $UW_2$. The segment d is parallel to the diagonal of the matrix and indicates the way in which the distance of any point of the straight lines to the sides of the matrix is measured.

**FIG. 2:** Example of an IPN for an arbitrary adjacency matrix of 1500 animal species (columns) and 500 plant species (rows) and $\phi = 0.2$. The shaded area with lines from top right to bottom left indicates the number of animal species with degree $p_o$ or more and the shaded area with lines top left to bottom right indicates the number of plant species with degree $a_o$ or more. A discrete stair-like approximation with steps of constant width $\Delta a$ is included in order to obtain the degree distribution of the animal species (columns).

**FIG. 3:** Experimental data of the two possible cumulative degree distributions for animal species (circles) and plant species (black triangles) for the system described in Robertson (1929), involving 1428 animals and 456 plants with $\phi = .023$. The continuous curve is the IPN obtained with the analytic expressions reported above. The endpoints of the cumulative degree distributions are indicated by arrows. Since both distributions are referred respectively to the horizontal and vertical axis, circles should be read from right to left, and triangles from top to bottom.

**FIG. 4:** Left panel: The empirical degree distributions for rows (open squares) and columns (filled triangles) for the adjacency matrix of the Robertson (1929) system, are shown together with the corresponding distributions derived from the IPN (rows: continuous line; columns: dashed line) for a matrix with the same dimensions and probability of contact $\phi$ than those empirically observed. Right panel: The same as in the left panel but with the data for columns renormalized using the procedure given in the text.

**Fig. 5.** Degree distributions of animal species of the Kato et al. (1990) system in three stages of the SNM with no FL (stars: after 1 iteration; triangles, after 10,000 iterations; open circles, after 100,000 iterations). Inset: same with an alternative swap criterion promoting interactions with lower-degree mutualist; All the distributions shown are the average over 100 realizations of the SNM.

**Fig 6.** Change of $T_m$, with the number of iterations of the SNM for the Kato et al. (1990) system. The letters indicate the corresponding adjacency matrices shown in the panels of Fig. 7.

**Fig 7.** Changes in the adjacency matrix of the pollination network of the Kato et al. (1990) system in several stages of the SNM. Contacts between species are shown as black pixels. Panel **A**, iteration 100 ($T_m$=9.3); panel **B**, iteration 1000 ($T_m$ =5.42); panel **C**, iteration 6665 ($T_m = T_r = 0.95$); panel **E**, iteration10,000 ($T_m$ =0.6) (the corresponding degree distribution is shown as triangles in Fig 5); panel **F,** iteration 100,000 ($T_m$ =0.02) (the corresponding degree distribution is shown as open circles in Fig 1). The experimentally observed distribution of contacts is shown in panel **D**.

**Fig 8.** Cumulative and normalized degree distributions of the animal mutualists of five real pollination networks (open circles). The corresponding distributions obtained with the SNM when $T_m \approx T_r$ are shown with filled squares. A better fit is obtained for the three larger systems when the SNM is run for a larger number of iterations and $T_m$ is allowed to drop below $T_r$ (theoretical distributions are shown with crosses). The number of iterations for each case is given in Table 1. The effect of introducing FL is shown in the panel for the Kato et al. (1990) system as a continuous line. In this case the degree distribution is the asymptotic one with a proportion of 80% of FL. In all cases the distributions that are shown are averages of 100 realizations.

FIGURE 1

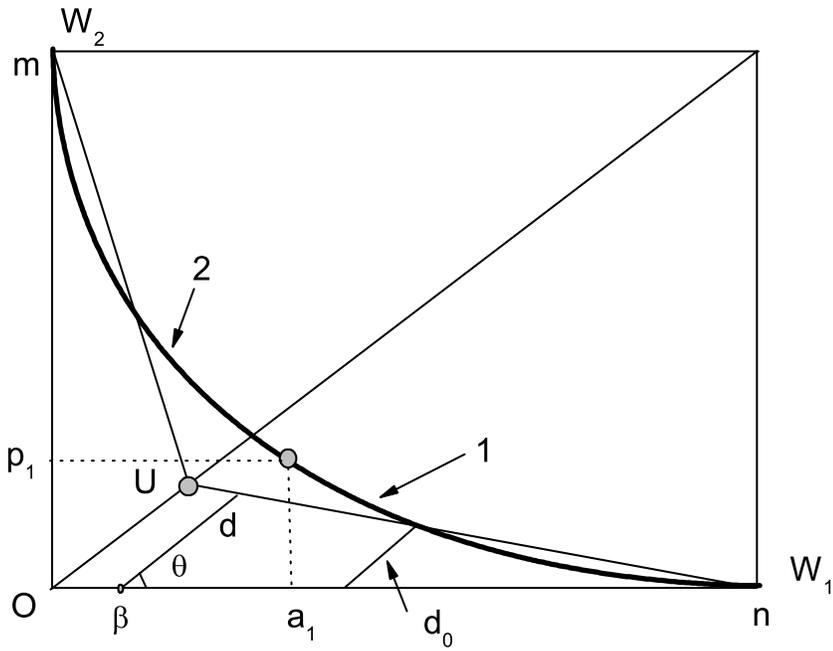

FIGURE 2

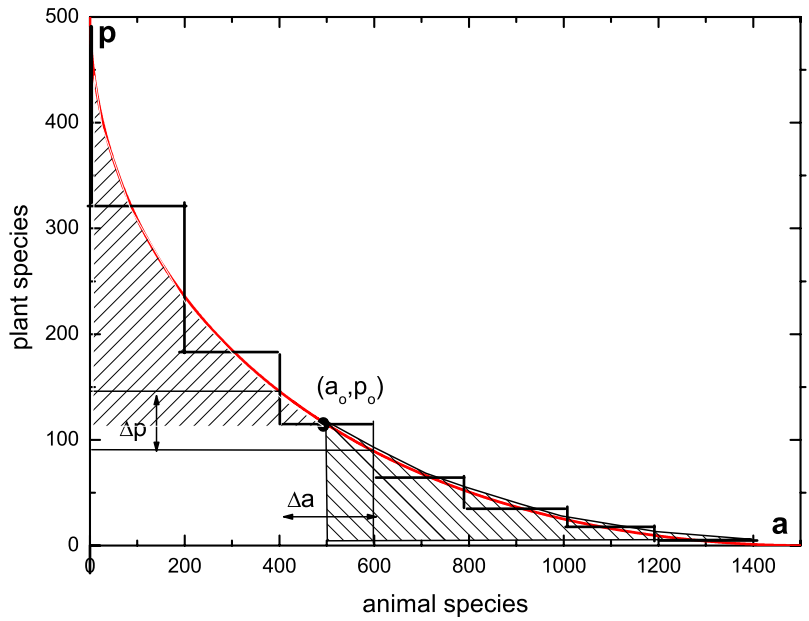

FIGURE 3

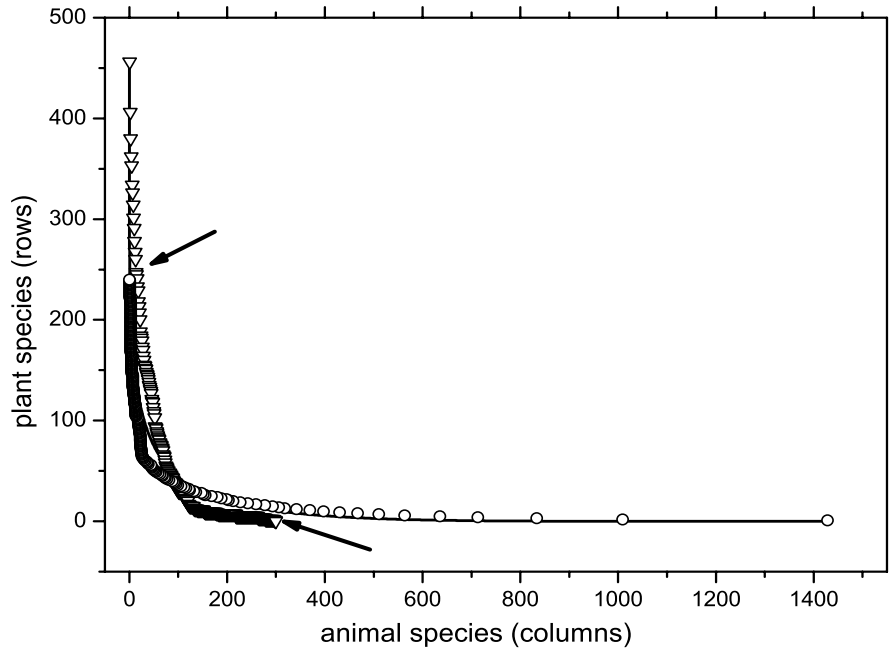

FIGURE 4

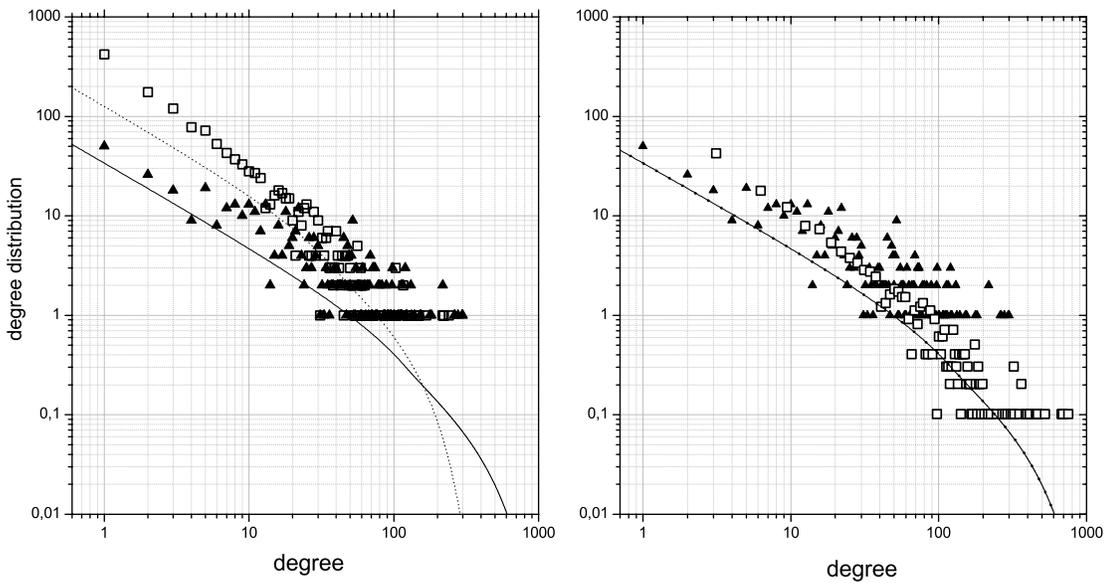

FIGURE 5

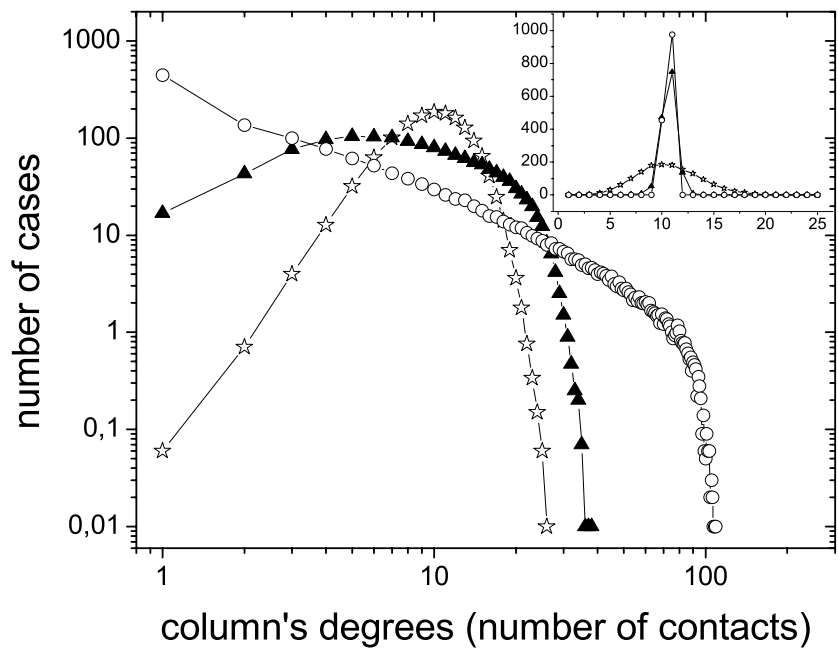

FIGURE 6

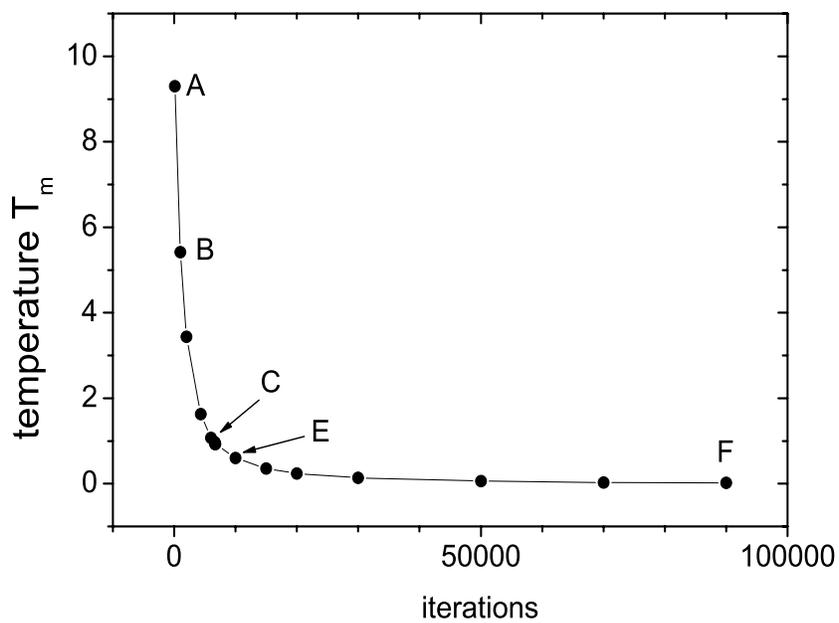

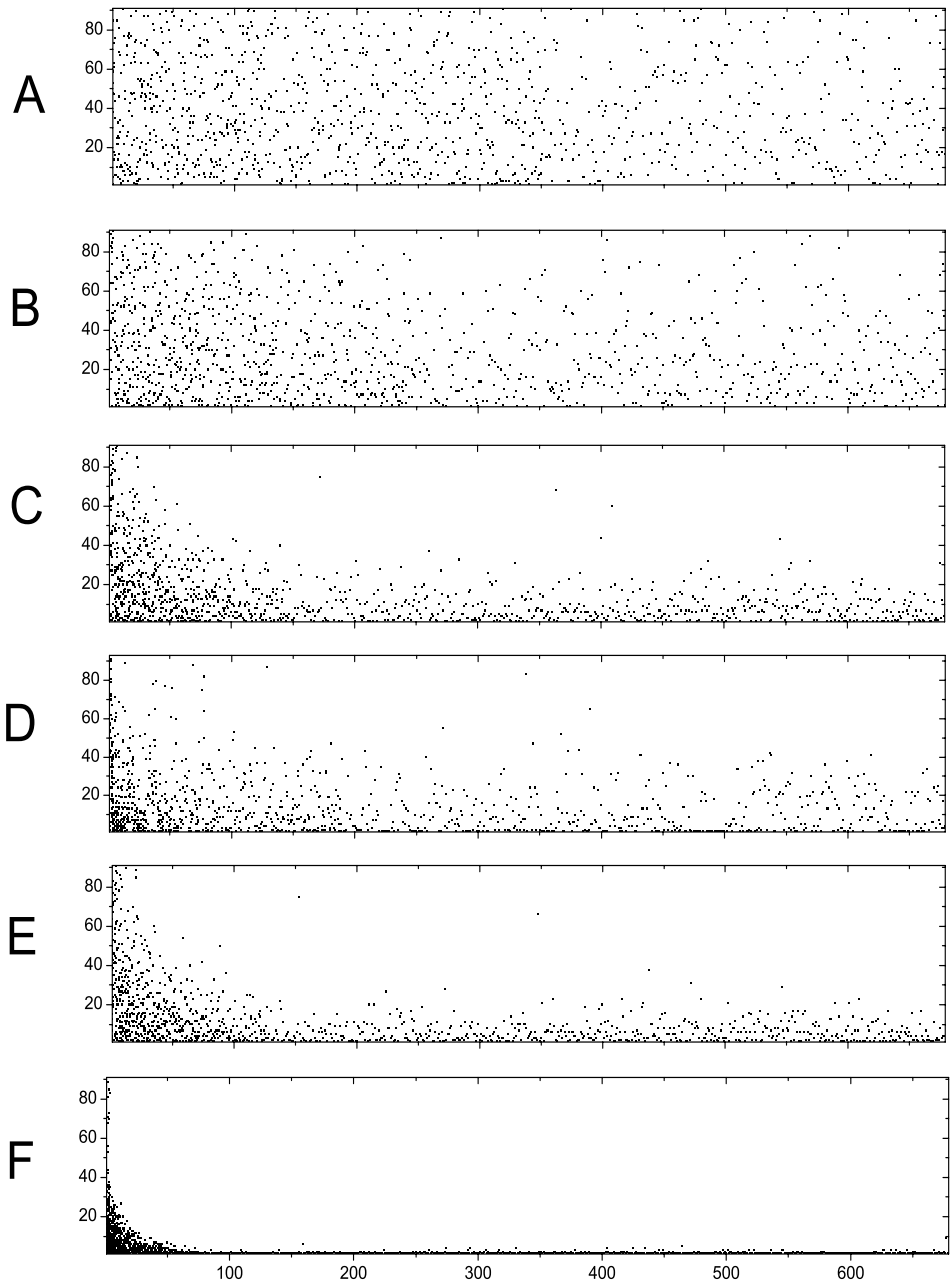

FIGURE 8

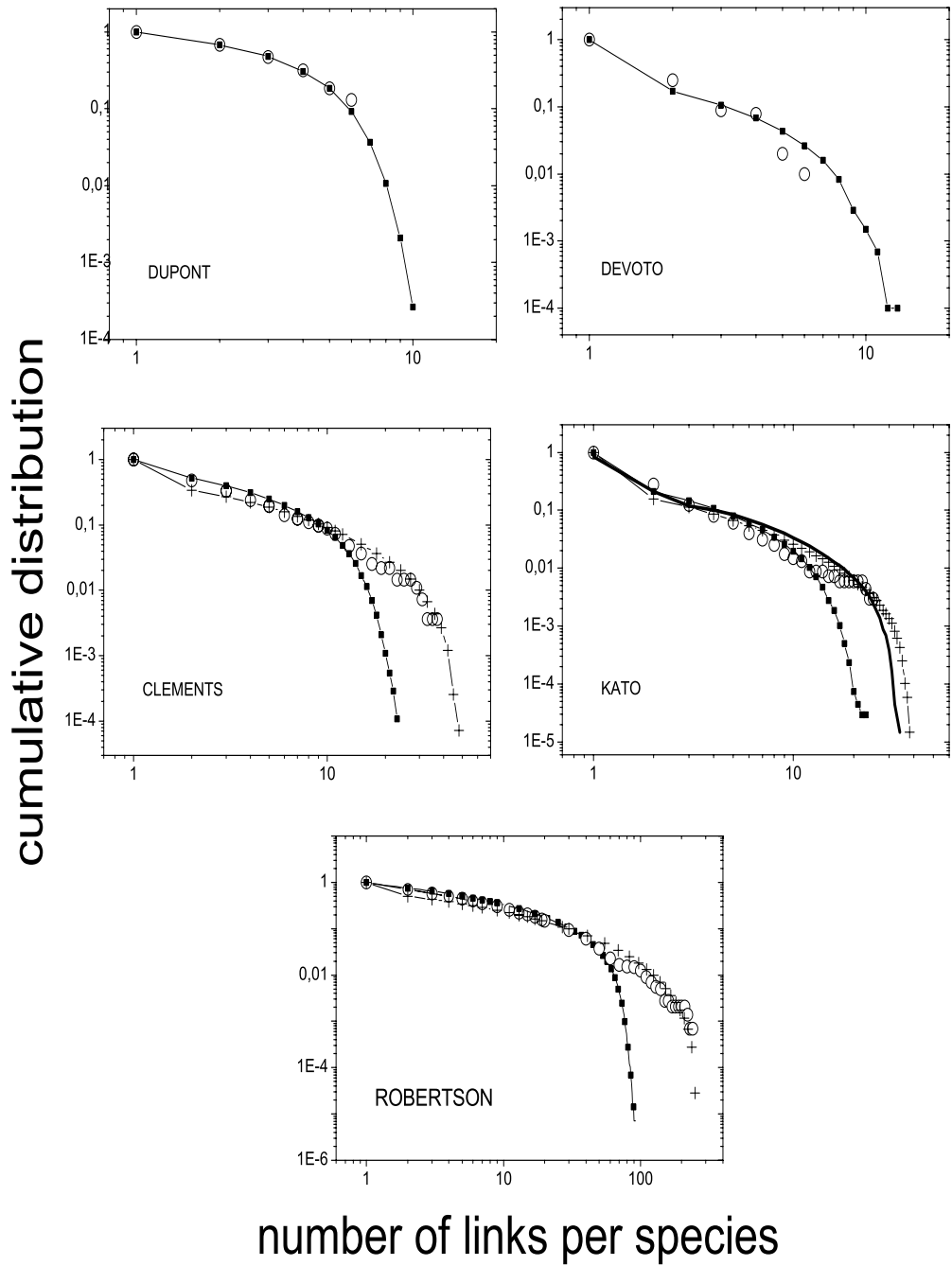